\documentclass[aps,prx,amsfonts,amsmath,amssymb,raggedbottom,longbibliography,reprint,superscriptaddress,citeautoscript]{revtex4-2}

\usepackage{graphicx} 
\usepackage{float}
\usepackage{color}
\usepackage{bm}
\usepackage{hyperref}
\usepackage{todonotes}
\usepackage{verbatim}
\usepackage{soul}
\usepackage{glossaries}
\usepackage{sidecap}
\usepackage{hyperref}
\usepackage{ulem}
\hypersetup{
    colorlinks=true,
    linkcolor=blue,
    filecolor=magenta,
    urlcolor=blue,
    citecolor=blue,
}

\begin{document}

\title{Superconductivity from the Slater mode: Application to KTaO$_3$ heterostructures}

\author{M. R. Norman}
\email{norman@anl.gov}
\affiliation{Materials Science Division, Argonne National Laboratory, Lemont, IL 60439}

\date{\today}

\begin{abstract}
Superconductivity has been observed for the 2D electron gas (2DEG) at the interface of KTaO$_3$ with other oxides, with a transition temperature about an order of magnitude higher than its $3d$ cousin SrTiO$_3$. The superconducting transition temperature is strongly dependent on the orientation of the interface. Motivated by this observation, we study pairing due to exchange of the soft transverse optic phonon mode characteristic of quantum paraelectrics and use the resulting theory to comment on the nature of superconductivity of this 2DEG. We find (1) an orientation dependence consistent with experiment along with an anisotropic gap function, but (2) a BCS coupling constant that is smaller than needed and so must be augmented by contributions from other phonons to be consistent with the observed values of T$_c$.
\end{abstract}

\maketitle

\section{Introduction}
The discovery of the 2D electron gas (2DEG) at the interface of the quantum paraelectric SrTiO$_3$ (STO) with other insulating oxides in 2004 \cite{ohtomo}, followed by the discovery of its superconductivity \cite{reyren}, renewed interest in superconductivity of low carrier density oxides that has been studied since the 1960s \cite{schooley}. One of the suggested mechanisms for this superconductivity is pairing via the soft transverse optical (Slater) phonon that is responsible for ferroelectricity \cite{edge}. Although the lowest order gradient coupling vanishes for low carrier densities (being quadratic in the atom displacements), a Rashba-like coupling is present at linear order in the displacements that involves hopping between $d$ orbitals of different symmetry. It is this mechanism that we explore in this paper.

Interest in this mechanism was renewed by the discovery of superconductivity in the related $5d$ quantum paraelectric KTaO$_3$ (KTO). To date, no superconductivity has been observed for bulk samples. For the 2DEG at the 001 surface, superconductivity at very low temperatures (less than 50 mK) was discovered by Iwasa's group using ionic liquid gating \cite{ueno}. This is where the field stood until Bhattacharya's group discovered superconductivity in excess of 2K for a 111 interface \cite{liu}. This was remarkable in that the value of T$_c$ was about an order of magnitude larger than what had been seen in STO, implying that the boost has something to do with spin-orbit coupling (the spin-orbit interaction of KTO is about twenty times larger than that of STO). This was followed by the discovery of superconductivity for the 110 interface at temperatures around half that for 111 \cite{chen}. Very recently, superconductivity was seen at a 001 interface with a T$_c$ of around one-tenth that observed for 111 by the use of a CaZrO$_3$ counterelectrode. In the most recent work, superconductivity was boosted at the 111 interface by Nb doping which drives the interface towards ferroelectricity \cite{eom}. These observations seem consistent with the soft TO1 (Slater) mode being a major driver of the superconductivity.

In previous work, a phenomenological theory based on TO1 mode pairing was proposed to not only explain the orientation dependence of T$_c$, but also the linear behavior of T$_c$ with carrier density \cite{liu2}. The basic idea behind this theory is that since the TO1 mode couples $t_{2g}$ electrons with different symmetry, then T$_c$ is sensitive to the orbital degeneracy.  For the 111 case, all three orbitals are degenerate (modulo small trigonal distortions, see Appendix A).  For 110, the $xy$ state is split to a higher energy via quantum confinement at the interface, whereas for 001, the $xz$ and $yz$ states are split off to a higher energy. Therefore, one can easily see that such a theory predicts that $T_c^{111} > T_c^{110} > T_c^{001}$ (since the $t_{2g}$ ground state degeneracy goes from 3 to 2 to 1), and this was confirmed by numerical calculations \cite{liu2}. Moreover, this theory predicts a divergence of T$_c$ as the TO1 mode energy goes completely soft at the quantum critical point for the appearance of ferroelectricity.  In the latter context, this theory has been used by another group to explain the recent results on Nb-doped KTO mentioned above \cite{yang-chen}.

In this paper, we put this phenomenological theory on a more solid footing by explicit calculations of the electron-phonon interaction based on previous work \cite{maria1} using recent ab-initio values derived for KTO \cite{giulia1}. The results for the 111 and 001 interfaces follow that of the phenomenological theory, but also provides new insights. In particular, the gap function is found to be strongly dependent on the energy band index as well as showing significant in-plane angular anisotropy.  Interestingly, this occurs for both the 111 and 001 cases.  On the other hand, the resulting BCS coupling constants are smaller than what one needs to explain the absolute magnitude of T$_c$. This indicates that other phonons must be involved in the pairing. This is consistent with recent work that shows a strong orientation dependence as well for the highest energy LO mode \cite{chen-feng}.

\section{Methods}
A microscopic theory for TO1 mode pairing has been developed by Gastiasoro {\it et al.} for bulk perovskites \cite{maria1} and this group has recently derived ab-initio values for KTO \cite{giulia1}. Here, we generalize this to the interface case. Based on previous work \cite{pablo,liu2,yang}, we will use a bilayer approximation for the electronic structure \cite{xiao} (Appendix A). This is motivated by the fact that such a simple model reproduces ARPES data \cite{bruno} and also aligns with the fact that the interface itself controls the superconductivity. The latter is known from the fact that the mean-field T$_c$ is boosted by negative gate voltages that act to push the electrons to the interface \cite{liu2} as previously observed for STO \cite{hwang}. In our previous work \cite{pablo,liu2}, we considered just the primary near-neighbor hopping ($xy$ to $xy$ along the $x$ and $y$ bonds, etc.) with the value of $t$ adjusted to fit the ARPES dispersion. This was motivated by the fact that such a model reproduces the pronounced star-shaped Fermi surface observed by ARPES for the 111 surface \cite{bruno} that has been recently seen as well for the 111 interface \cite{UBC}. We then contrast those results with a tight binding model based on ab initio DFT which involves as well the secondary near-neighbor hopping ($xy$ to $xy$ along $z$, etc.), as well as two next-near-neighbor hoppings (one diagonal in the orbital index, the other off-diagonal) \cite{xiao}. This model was used in recent work we have done on the 110 interface \cite{yang} with values extracted from the Materials Project \cite{MP}.  These hopping values, as well as the spin-orbit coupling, are listed in Table \ref{table1}. We also include a small static Rashba term \cite{pablo} because of the broken inversion symmetry of the interface. This is useful as well as it lifts the Kramers degeneracy at each $k$-point and so allows us to investigate the helicity dependence of the superconducting order parameter. By helictty, we mean the two Rashba-split Fermi surfaces (as an example, the red and blue curves in Fig.~1a) that are related at each $k$ by an approximate spin flip.

\begin{table}
\caption{The tight binding parameters for KTO used in the present work. TB model 1 (TB1) is taken from Ref.~\cite{pablo}. TB model 2 (TB2) is based on bulk DFT values as discussed in Ref.~\cite{yang}. $t$ is the primary nearest-neighbor hopping ($xy$ to $xy$ along $x,y$, etc.), $t'$ the secondary nearest-neighbor hopping ($xy$ to $xy$ along $z$, etc.), $t''$ the next-nearest-neighbor hopping diagonal in the orbital index, and $t'''$ the next-nearest-neighbor hopping that is off-diagonal in the orbital index. Details for the 111 bilayer case are given in Ref.~\cite{xiao}. $\xi_{so}$ is the spin-orbit coupling, and $\mu$ the chemical potential relative to the band 1 energy at $\Gamma$. All values are in eV.}
\begin{ruledtabular}
\begin{tabular}{lll}
 & TB1 & TB2\\
\hline
$t$ & 1.0 & 0.4975\\
$t'$ & 0 & 0.035\\
$t''$ & 0 & 0.09\\
$t'''$ & 0 & 0.0175\\
$\xi_{so}$ & 0.265 & 0.265\\
$\mu_{111}$ & 0.0775 & 0.11565\\
$\mu_{001}$ & 0.4884 & 0.3642\\
\end{tabular}
\end{ruledtabular}
\label{table1}
\end{table}

The electron-phonon coupling constants due to the TO1 mode can by thought of as dynamic Rashba terms. In Ref.~\cite{giulia1}, the dynamic Rashba hoppings were derived for KTO. The dominant term is just the dynamic generalization of the static term mentioned above which involves inversion-breaking hopping (generated primarily by the TO1 mode oxygen vibrations normal to the metal-oxygen bonds) within the $t_{2g}$ manifold of states:
\begin{equation}
t_{2g,i} \rightarrow 2p_j \rightarrow t_{2g,k}
\end{equation}
where $\rightarrow$ indicates a hop. Here, $i,k$ run over $yz,xz,xy$ and $j$ over $x,y,z$. This term is non-zero for non-equal $t_{2g}$ orbital indices if the oxygen ion is displaced off the bond and is independent of spin so involves the Pauli $\sigma_0$ operator \cite{shanavas}. But they also derived the three spin-dependent dynamic hoppings (proportional to the Pauli spin matrices $\sigma_i$). These are almost certainly due to virtual hoppings between the $t_{2g}$ and unoccupied $e_g$ orbitals \cite{shanavas,shanavas2,kim}:
\begin{equation}
t_{2g,i} \rightarrow 2p_j \rightarrow e_{g,m} \sim t_{2g,k}
\end{equation}
where $\sim$ indicates a spin-orbit transition and $m$ runs over $x^2-y^2$ and $3z^2-r^2$. To appreciate this, we list the values of these ab-initio terms in Table \ref{table2}. Had these spin-dependent hoppings been due to the spin-orbit matrix elements between different oxygen $2p$ orbitals, then (1) the magnitude of the three terms would be identical and (2) the values would be small since the spin-orbit coupling of the $2p$ electrons is small. Instead, if they were due to hoppings between the $t_{2g}$ and $e_g$ orbitals, one can easily show that the term $t_B$ in Table \ref{table2} should be small (as Table \ref{table2} indeed indicates) and also that these terms are far more important in KTO than in STO. The latter arises from the spin-orbit coupling of the $d$ electrons being 20 times larger, and the value of these terms by lowest order perturbation theory goes as $t_d \xi_{so}/\Delta_{CF}$ where $t_d$ is the dynamic hopping between the $t_{2g}$ and $e_g$ orbitals, $\xi_{so}$ is the spin-orbit coupling connecting the $t_{2g}$ and $e_g$ orbitals, and $\Delta_{CF}$ is the splitting between the two sets of orbitals (approximately 4 eV) \cite{shanavas,shanavas2}, with the spin-dependence due to $\xi_{so}$. This has been verified by ab initio calculations \cite{kim}. Regardless of their origin, we consider all four dynamic Rashba terms in the work below.  The primary difference from the previous phenomenological work is that the value of $t_0$ in Table \ref{table2} is about 50\% larger than the value assumed in Ref.~\cite{liu2}. Since the BCS coupling constant $\lambda$ goes as the square of $t_0$, this provides a large boost compared to before. The contributions to $\lambda$ from $t_i$ are more modest, but are significant as well.

\begin{table}
\caption{Dynamic Rashba terms from Ref.~\cite{maria1} as in Appendix A. $t_0$ is in eV, and the rest are in units of $t_0$.
The middle column is for the static case (Fig.~1a), the right column for the dynamic case (Fig.~1b).
For the 111 dynamic case, these values are reduced by $1/\sqrt{3}$ as discussed in the text.}
\begin{ruledtabular}
\begin{tabular}{lll}
$t_0$ & 0.002 & 0.0488\\
$t_A$ & 1.08 & 1.08\\
$t_B$ & 0.1 & 0.1\\
$t_C$ & 0.44 & 0.44\\
\end{tabular}
\end{ruledtabular}
\label{table2}
\end{table}

In the Cooper channel, one has a ladder sum involving scattering of the the Cooper pair $nk,-nk$ to the pair $n'k',-n'k'$ where $n,n'$ are band indices. Based on Ref.~\cite{maria1}, the secular matrix for the linearized gap equation  can be written as:
\begin{eqnarray}
A_{nk,n'k'} & = & c N_n N_{n'} wt_k wt_{k'} \sum_{\alpha \alpha'} |<n\alpha k|V_R|n'\alpha' k'>|^2 \nonumber \\
& & [2\omega_0/\omega^2(q)] – N_n wt_k \delta_{nk,n'k'}
\label{matrix}
\end{eqnarray}
where $c$ is formally equal to $\ln(1.14 \omega_c /T_c)$ and thus $1/\lambda$ with $\omega_c$ the BCS cut-off.  Here, $n,n'$ run from 1 to 4 (i.e., two pairs of bands, each pair being split by the static Rashba term), $N_n$ is the density of states for band $n$, $w_k$ are the k-point weights ($k$ refers to the irreducible wedge of the zone, with $\sum_kwt_k=1$), $\alpha,\alpha'$ are the group operations (12 for 111, 8 for 001), where the sum in Eq.~\ref{matrix} has been normalized by the square of the number of group operations. $V_R$ is the dynamic Rashba interaction (a sum over the $t$ operators in Table \ref{table2}, with the matrix elements given in Appendix A), and $\omega_q$ is the TO1 mode energy at a transferred momentum $q=k'-k$ (with the band index implicit in $k$). Note one power of $\omega_q$ is coming from the electron-phonon matrix elements, the other power from the phonon propagator (the factor of 2 in Eq.~\ref{matrix} comes from the phonon propagator; we assume as in Ref.~\cite{maria1} a static approximation where the propagator is $-2/\omega_q$ \cite{mcmillan}). Here, $\omega_0$ is the value of the TO1 mode that was assumed when deriving the values in Table \ref{table2} (2.5 meV, the undoped value for bulk KTO) and hence must be backed out when allowing $\omega(0)$ to deviate from this value (which is why it shows up in the numerator). As in all weak coupling calculations, we assume $k,k'$ are confined to the Fermi surface. In our numerical simulations, we use an angular step factor of $1^\circ$ for the Fermi surface angle, and thus the size of the secular matrix is 124 for the 111 case ($30^\circ$ wedge) and 184 for the 001 case ($45^\circ$ wedge).

An important point from Eq.~\ref{matrix} is that the dynamic Rashba terms are evaluated at $k_0=(k+k')/2$ in the Cooper channel since one is scattering at the electron-phonon vertex from $k=k_0-q/2$ to $k'=k_0+q/2$. One can easily see that this leads to a strong peak in the interaction for forward scattering ($k'=k$) and a complete suppression for back scattering ($k'=-k$) as the Rashba interaction scales linearly with $k_0$. This point is evident from the simplified bulk calculation presented in Ref.~\cite{maria1}.

When evaluating the $V_R$ matrix elements, we assume only coupling to the TO1 mode that is polarized perpendicular to the interface (which also underlies the static Rasbha splitting at the interface). The reason we neglect coupling to the other mode can be appreciated from the 001 case.  For the $t_0$ terms in Table \ref{table2}, this mode in the 001 case only couples $yz$ to $xz$ orbitals, and the coupling \cite{maria1} goes as $\vec{k_0}\cdot \vec{q}$. In a simple approximation where the Fermi surface is circular, this vanishes in that $\vec{k_0}$ is transverse to $\vec{q}$.  So, we expect in general that coupling to this mode is small and so we ignore it, though a more complete calculation than what we present here would take it into account.

Finally, the form of Eq.~\ref{matrix} implicitly assumes that $\Delta_{nk}$ (the BCS order parameter) is invariant under group operations; that is, we assume it comes from the identity representation (i.e., it is invariant under $\alpha$).  We expect this to have the highest $T_c$ given the attractive nature of the electron-phonon interaction. Note that the appearance of the extra factors of $N_n$ and $w_k$ was designed to make $A$ a symmetric matrix so that standard eigen routines can be used. $c$ is then derived by finding when the first eigenvalue of $A$ crosses zero. The resulting eigenvector is $\Delta_{nk}$ (with relative values since this is the linearized gap equation).

\section{Results}
We start with the 111 case.  An important point is that the out-of-plane phonon polarization is $n_{ph}=(1,1,1)/\sqrt{3}$, and so the values of $t_i$ in Table \ref{table2} need to be scaled down by a factor of $1/\sqrt{3}$ when the matrix elements in Eq.~\ref{matrix} are calculated for the 111 case. We first consider the Fermi surface in our bilayer approximation \cite{xiao} where the chemical potential has been adjusted to give a 2D carrier density of $n_{2D}=10^{14}cm^{-2}$.  For now, we consider the tight binding parameters from the middle column of Table \ref{table1} that we used in our previous work \cite{pablo,liu2} (TB model 1). In Fig.~1a, we show the Fermi surface with a small static Rashba coupling of 2 meV.  This small term was added to lift Kramers degeneracy and will be used as the basis for evaluating Eq.~\ref{matrix}.  We compare this plot to the `dynamic' case (Fig.~1b) that use the (scaled) values from the right column of Table \ref{table2}. What this plot represents is the maximum distortion of the Fermi surface during a dynamic (TO1) oscillation. From Fig.~1a, we note the pronounced star shape of the outer Fermi surfaces. This resembles ARPES data \cite{bruno,UBC} which was the original motivation for this tight binding model \cite{pablo}.  From Fig.~1b, one sees the profound impact the dynamic Rashba splitting has on the Fermi surface, noting that the chemical potential was left to its `static' value from Table \ref{table1}. This demonstrates the significant impact the dynamic Rashba coupling has on the electronic structure. In the context of Eq.~\ref{matrix}, these plots represent the `forward scattering' limit ($n=n'$, $k=k'$).

\begin{figure}
\centering
\includegraphics[width=1.0\columnwidth]{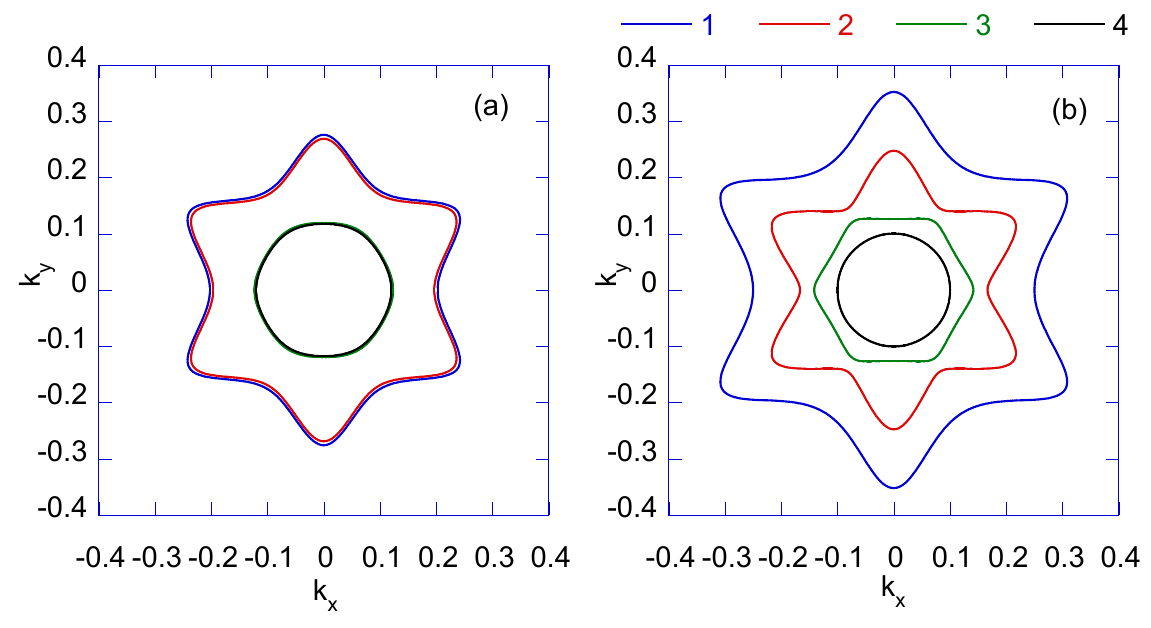}
\caption{Fermi surface for the bilayer 111 tight binding model (TB model 1) with $\mu$ set to give $n_{2D}=10^{14}cm^{-2}$. (a) the static Rashba case ($t_0=2$ meV) and (b) the dynamic Rashba case ($t_0=48.8/\sqrt{3}$ meV).  $k$ is in units of $\pi/c$ where $c=\sqrt{2/3}a$ with $a$ the bulk lattice constant.  $k_x$ ($\Gamma-K$) is along the (1,-1,0) direction and $k_y$ ($\Gamma-M$) along the (-1,-1,2) direction. Note the profound impact of the dynamic Rashba coupling on the Fermi surface.}
\label{fig1}
\end{figure}

We now investigate the electron-phonon kernel. To do this, we use the $nk$ eigenvectors from the results of Fig.~1a. From Eq.~\ref{matrix}, we plot
\begin{equation}
|<nk|V_R|nk'>|^2 [2\omega_0/\omega^2(q)]
\label{vr}
\end{equation}
in Fig.~2 for two representative energy bands (one of the outer Fermi surfaces, band 1, and one of the inner Fermi surfaces, band 3). To do this, we need a model for $\omega_q$.  As in Ref.~\cite{liu2}, we use a Vaks parameterization of the TO1 mode dispersion for KTO that was derived from inelastic neutron scattering data at 10 K \cite{farhi}. This comes from solving a $2 \times 2$ matrix involving the TO1 mode and the acoustic TA mode (for simplicity, we ignore the small angular anisotropy terms). This dispersion ($q=k'-k$) is plotted around the Fermi surface for the outer Fermi surface (band 1) in Fig.~3 as a function of $\phi'-\phi$ for two different values of $k$: $\phi$=0 is along the $\Gamma-K$ ($k_x$) direction in Fig.~1 and $\phi$=30 is along the $\Gamma-M$ direction. Here $\phi'-\phi=0$ corresponds to forward scattering.  For these plots, we assume $\omega(0)$ has hardened from 2.5 meV (bulk value) to 5.6 meV due to carrier screening at this value of the electron filling as estimated in Ref.~\cite{liu2} from electric-field dependent Raman data \cite{fleury} (Appendix B). As expected, $\omega(q)$ peaks for backward scattering since $q$ for this case is maximal (2$k_F$). What one observes in Fig.~2 is exactly the effect mentioned above, where the kernel peaks for forward scattering ($\phi'-\phi=0$) and is strongly suppressed for back scattering ($\phi'-\phi=180$). There are two contributions to this, one coming from $V_R$ which has a tendency to peak either at or near forward scattering (top plots), and the additional contribution in  brackets from the phonon dispersion where $\omega(q)$ has its lowest energy for forward scattering and largest for back scattering (the bottom plots include both contributions).

\begin{figure}
\centering
\includegraphics[width=1.0\columnwidth]{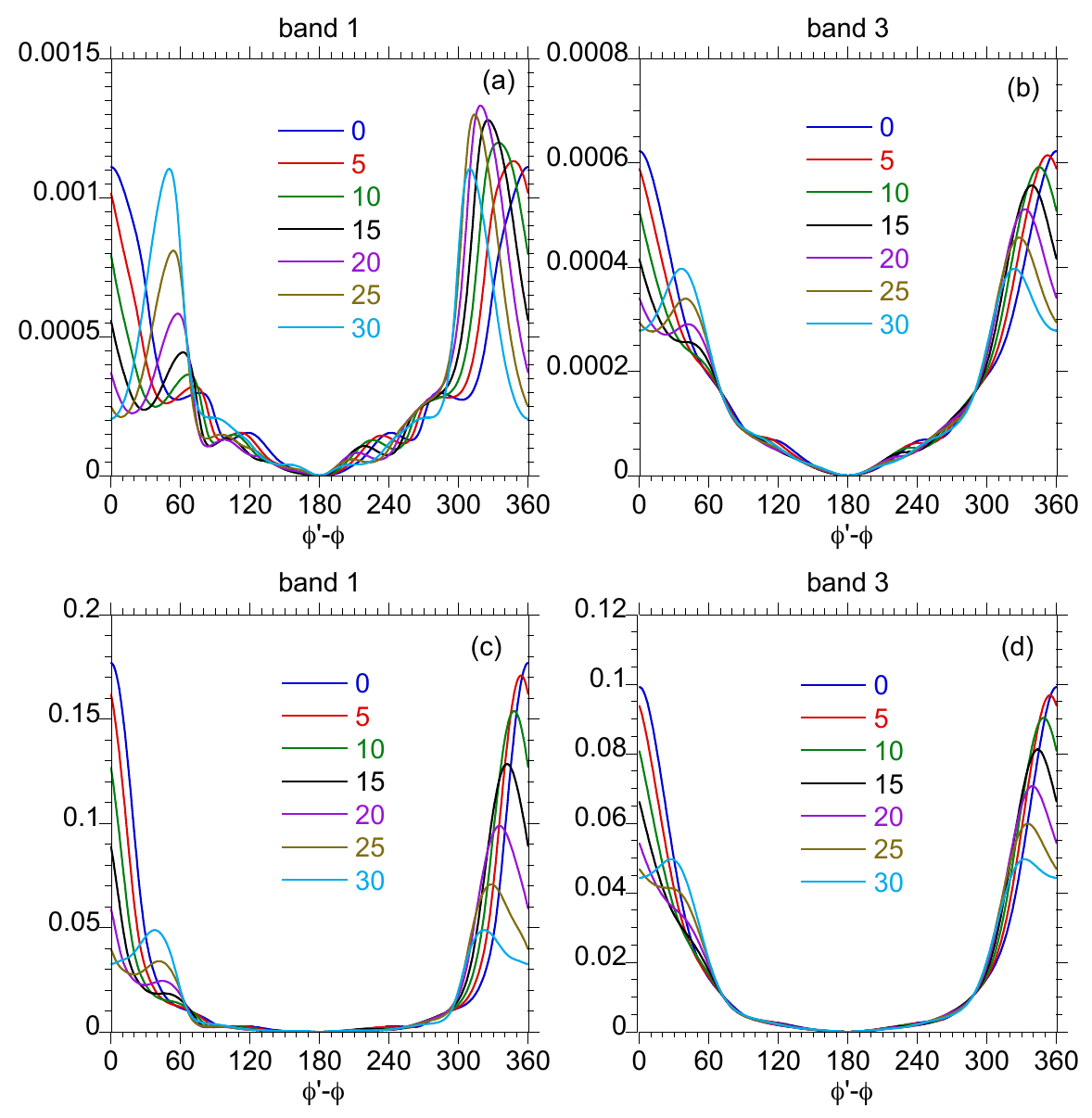}
\caption{Plots of Eq.~\ref{vr} using the Fermi surface from Fig.~1a versus $\phi'-\phi$ (where $\phi$ is the Fermi surface angle for $k$ and $\phi'$ for $k'$) for various $\phi$.  $\phi'-\phi=0$ corresponds to forward scattering ($k'=k$). Plots are for one of the outer Fermi surfaces (band 1) and one of the inner Fermi surfaces (band 3). The top plots do not include the term in brackets in Eq.~\ref{vr} (the phonon dispersion), the bottom plots do.  Note the tendency for the interaction to peak near forward scattering and to be completely suppressed for back scattering. This tendency is even more pronounced when the phonon dispersion is included.}
\label{fig2}
\end{figure}

\begin{figure}
\centering
\includegraphics[width=0.5\columnwidth]{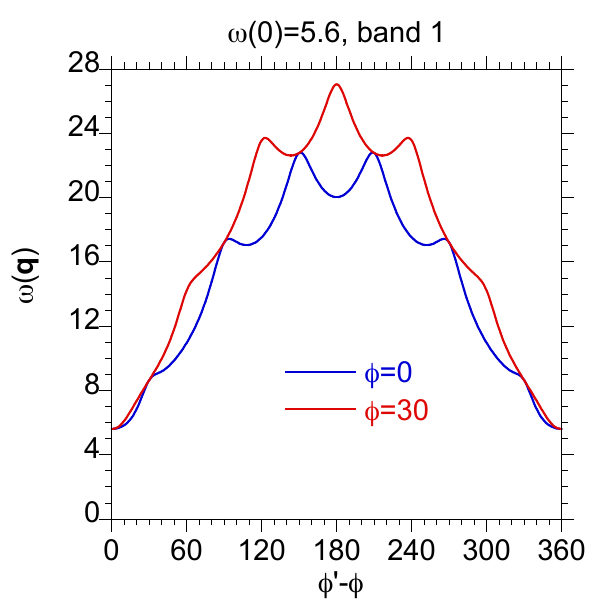}
\caption{Phonon dispersion $\omega(q=k'-k)$ as a function of $\phi'-\phi$ for the outer Fermi surface (band 1) from Fig.~1a for two different values of $\phi$ assuming $\omega(0)$=5.6 meV.  $\phi'-\phi=0$ corresponds to $q=0$ and 180 corresponds to $q=2k_F$. The larger value for $\omega(q)$ at $\phi'-\phi=180$ for $\phi=30$ is due to being at the tip of the star in Fig.~1a.}
\label{fig3}
\end{figure}

With these results, we now show the electron-phonon kernel from Eq.~\ref{matrix}
\begin{equation}
\sum_{k'} |<n k|V_R|n'k'>|^2 [2\omega_0/\omega^2(q)]
\label{gto}
\end{equation}
in Fig.~4 as a function of the Fermi surface angle for $k$, with the intraband terms ($n=n'$) plotted in (a) and the interband terms in (b) and (c). Here, the sum is over all $k'$ of a given Fermi surface ($n$).  One can think of this intermediate result as the contribution to the gap equation if the gap function had no dependence on $nk$.  Note the intraband terms all have similar behavior, but the interband terms differ. The largest ones connect one of the outer band helicity surfaces to its inner band counterpart. That is, band 1 to band 3 ($nn'$=13, 31) and band 2 to band 4 ($nn'$=24, 42).

\begin{figure}
\centering
\includegraphics[width=1.0\columnwidth]{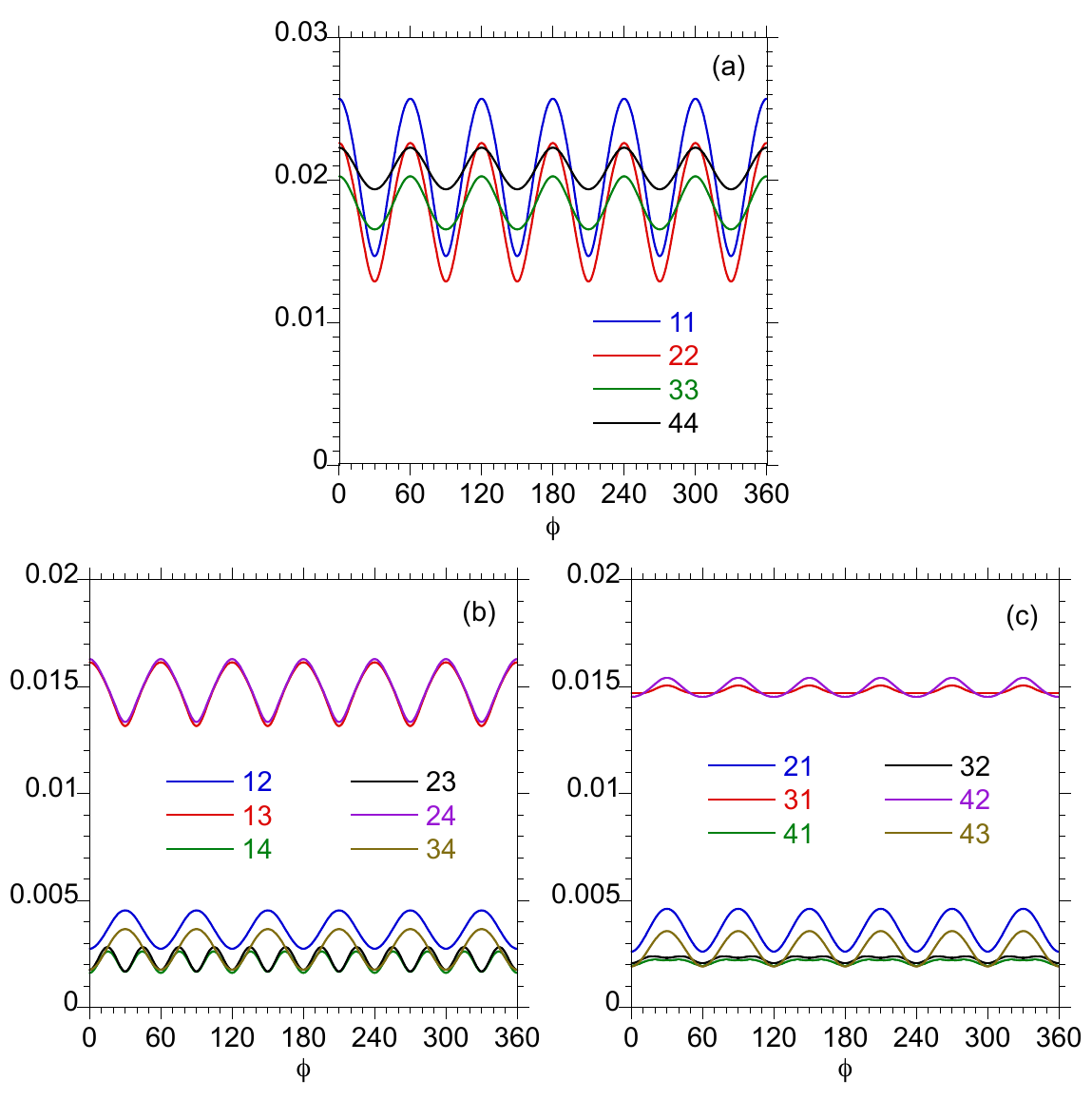}
\caption{Plots of Eq.~\ref{gto} versus $\phi$ (the Fermi surface angle for $k$) using the Fermi surface from Fig.~1a for various $n,n'$. (a) are the intraband terms ($n=n'$) and (b) and (c) the interband terms. The largest interband terms connect bands with the same helicity ($nn'=13, 31, 24, 42$).}
\label{fig4}
\end{figure}

In Fig.~5a, we show the solutions of the gap equation (Eq.~\ref{matrix}) in the irreducible wedge of the zone for the four bands.  Fig.~5b shows what would happen if we ignore the phonon dispersion; that is, set $\omega(q)=\omega(0)$, which is the approximation used in Ref.~\cite{maria1}.  Note that $\Delta_{nk}$ has a significant variation with band index and also (for some bands) a significant angular anisotropy. Also, the angular and band index dependence differs in the two cases. By definition, $\lambda$ is much larger in (b) since this plot assumes $\omega(q)=\omega(0)$.

\begin{figure}
\centering
\includegraphics[width=1.0\columnwidth]{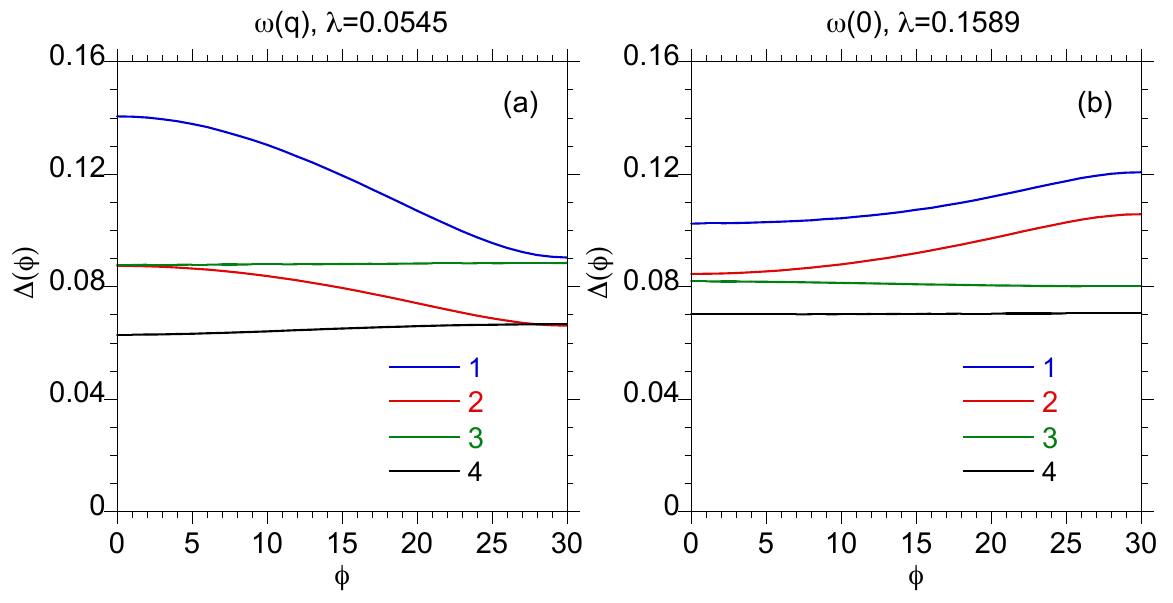}
\caption{The superconducting order parameter $\Delta_{nk}$ versus $\phi$ (the Fermi surface angle for $k$) using the Fermi surface from Fig.~1a for the different bands. (a) includes the phonon dispersion, (b) does not, i.e., $\omega(q)=\omega(0)$. The plot headers list the value of the BCS coupling constant, $\lambda$.}
\label{fig5}
\end{figure}

We now study the sensitivity of these results to the electronic structure.  Fig.~6 is a repeat of Fig.~1, but now using the tight binding parameters from the right column of Table \ref{table1} (TB model 2). In comparison to Fig.~1, the outer band Fermi surfaces are more rounded. Fig.~7 is a repeat of Fig.~5 for this electronic structure. Note the angular dependence and band index dependence of $\Delta_{nk}$ has changed, and $\lambda$ has a modest suppression, relative to Fig.~5.  So, as expected, the results have some sensitivity to the assumed electronic structure.

\begin{figure}
\centering
\includegraphics[width=1.0\columnwidth]{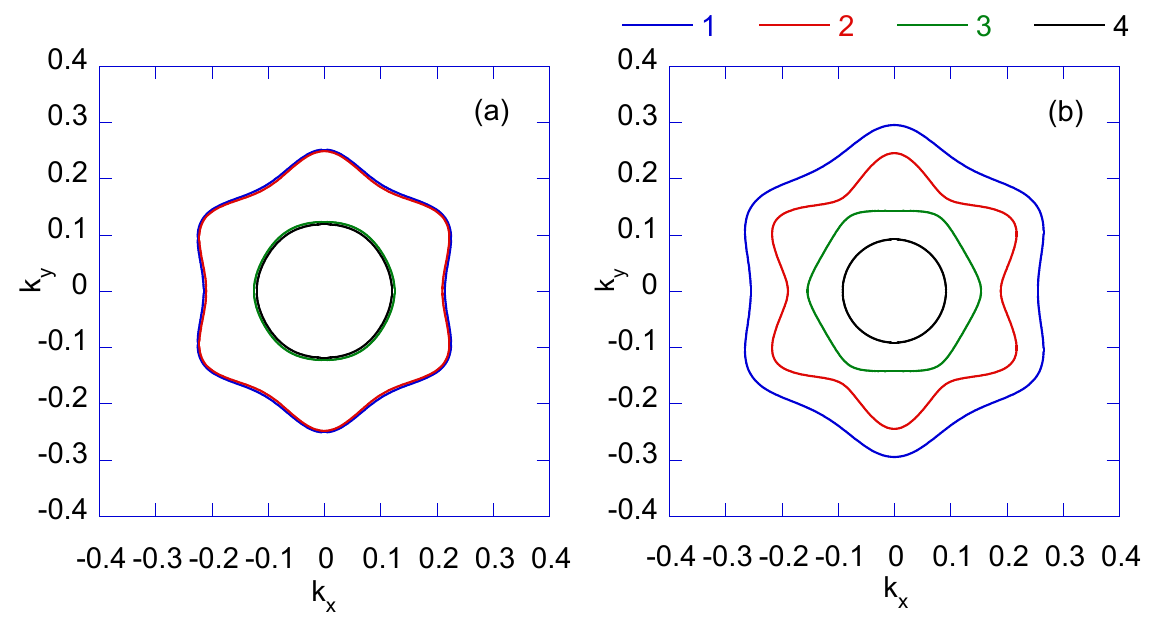}
\caption{Fermi surface for the bilayer 111 tight binding model (TB model 2) with $\mu$ set to give $n_{2D}=10^{14}cm^{-2}$. (a) the static Rashba case ($t_0=2$ meV) and (b) the dynamic Rashba case ($t_0=48.8/\sqrt{3}$ meV).  $k$ is in units of $\pi/c$ where $c=\sqrt{2/3}a$ with $a$ the bulk lattice constant.  $k_x$ ($\Gamma-K$) is along the (1,-1,0) direction and $k_y$ ($\Gamma-M$) along the (-1,-1,2) direction. Note the profound impact of the dynamic Rashba coupling on the Fermi surface, and the differences from Fig.~1 that used TB model 1 instead.}
\label{fig6}
\end{figure}

\begin{figure}
\centering
\includegraphics[width=1.0\columnwidth]{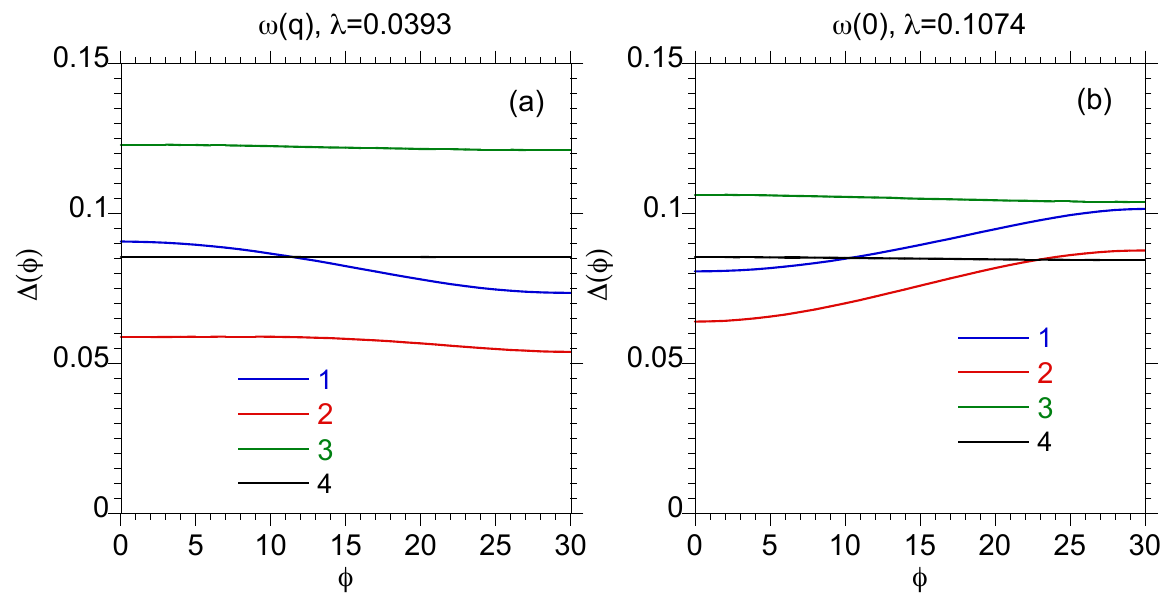}
\caption{The superconducting order parameter $\Delta_{nk}$ versus $\phi$ (the Fermi surface angle for $k$) using the Fermi surface from Fig.~6a (TB model 2) for the different bands. (a) includes the phonon dispersion, (b) does not. i.e., $\omega(q)=\omega(0)$. The plot headers list the value of the BCS coupling constant, $\lambda$. Note the differences from Fig.~5 that used TB model 1 instead.}
\label{fig7}
\end{figure}

We now turn to the 001 case. We did calculations for both sets of tight binding parameters, but for brevity, we just show the results from TB model 2 (though values for $\lambda$ for both models are shown in Table \ref{table3}). Figs.~8-10 are the 001 results (corresponding to Figs.~1, 4 and 5). Surprisingly, a significant angular anisotropy of the dynamic Rashba splitting is seen in Fig.~8b. Unlike for the 111 case, the only significant interband terms in Fig.~9 are the ones that connect bands 1 and 2 (and so, have opposite helicity). The larger splitting off of the band 1 Fermi surface relative to the other three in Fig.~8 is reflected in the angular anisotropy of $\Delta_{nk}$ in Fig.~10 (the angular dependence of $\Delta_{1k}$ roughly follows the dynamic Rashba splitting of the outer Fermi surfaces). As expected, $\lambda$ for the 001 case is significantly smaller than for the 111 case. Similar results are found using TB model 1.

\begin{figure}
\centering
\includegraphics[width=1.0\columnwidth]{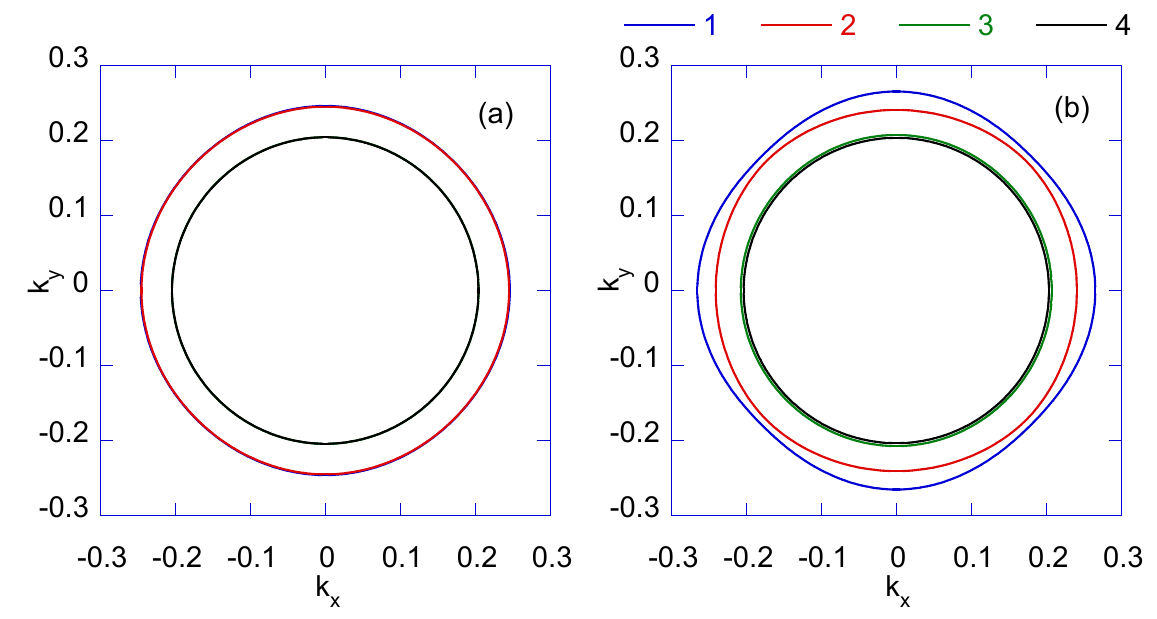}
\caption{Fermi surface for the bilayer 001 tight binding model (TB model 2) with $\mu$ set to give $n_{2D}=10^{14}cm^{-2}$. (a) the static Rashba case ($t_0=2$ meV) and (b) the dynamic Rashba case ($t_0=48.8$ meV).  $k$ is in units of $\pi/a$ where $a$ is the bulk lattice constant.  $k_x$ is along the (100) direction and $k_y$ along the (010) direction. Note the impact of the dynamic Rashba coupling, particularly for the outer (band 1) Fermi surface.}
\label{fig8}
\end{figure}

\begin{figure}
\centering
\includegraphics[width=1.0\columnwidth]{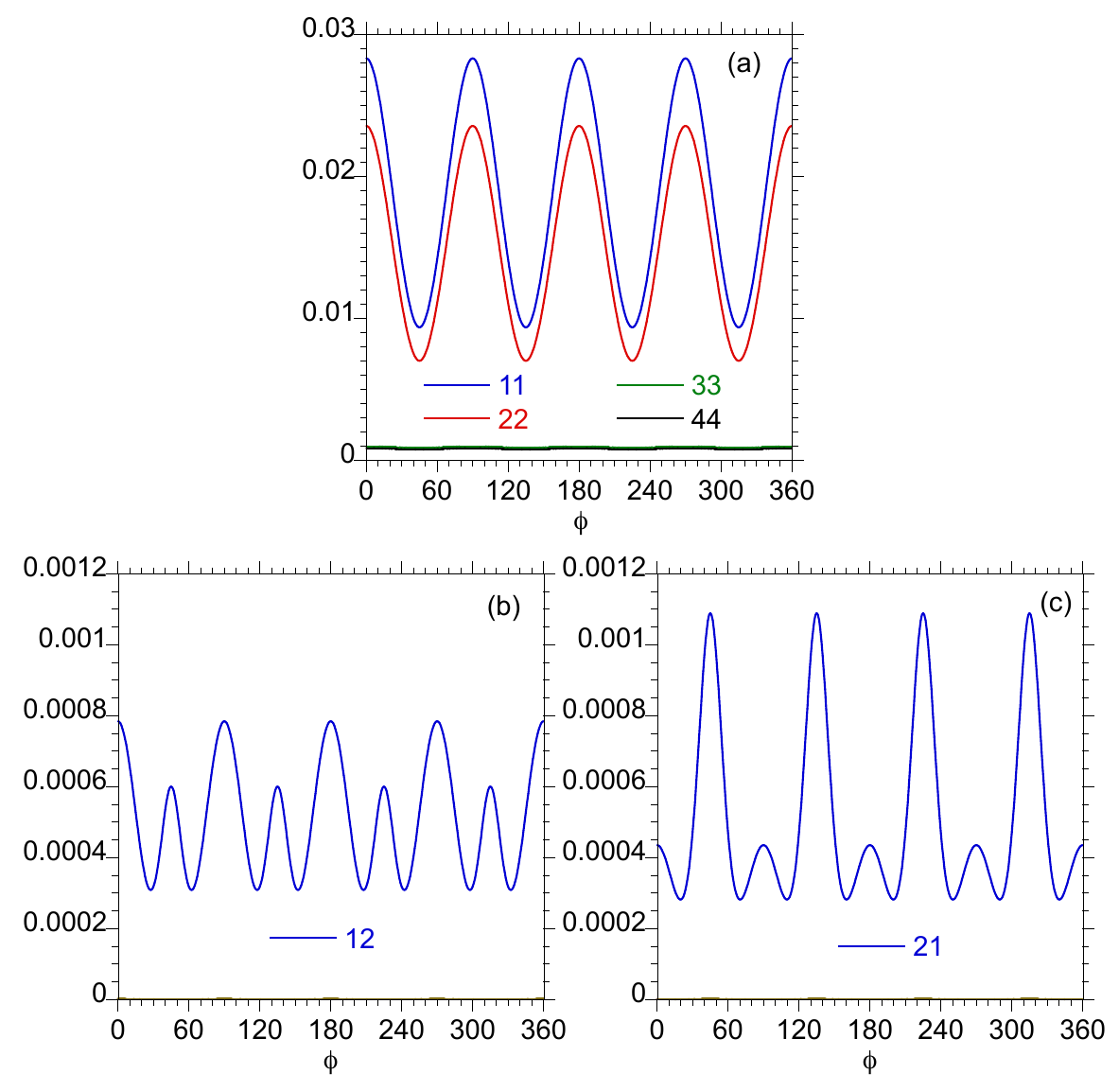}
\caption{Plots of Eq.~\ref{gto} versus $\phi$ (the Fermi surface angle for $k$) using the Fermi surface from Fig.~8a for various $n,n'$. (a) are the intraband terms ($n=n'$) and (b) and (c) the interband terms. Unlike for the 111 case, the only significant interband terms are the ones that connect the two outer Fermi surfaces ($nn'=12, 21$).}
\label{fig9}
\end{figure}

\begin{figure}
\centering
\includegraphics[width=1.0\columnwidth]{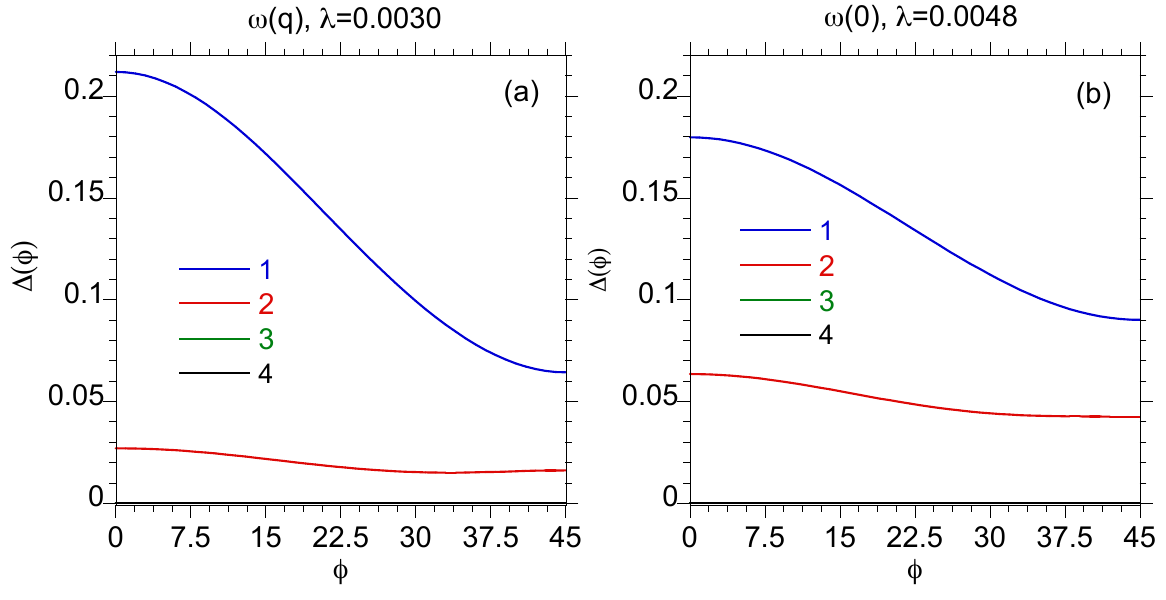}
\caption{The superconducting order parameter $\Delta_{nk}$ versus $\phi$ (the Fermi surface angle for $k$) using the Fermi surface from Fig.~8a (TB model 2) for the different bands. (a) includes the phonon dispersion, (b) does not, i.e., $\omega(q)=\omega(0)$. The plot headers list the value of the BCS coupling constant, $\lambda$. Note the significant angle dependence of $\Delta_{1k}$ that roughly follows the dynamic Rashba splitting of the outer Fermi surfaces shown in Fig.~8b.}
\label{fig10}
\end{figure}

In Table \ref{table3}, we summarize the results for $\lambda$ for the 111 and 001 cases: (1) for both tight binding models, (2) for two values of $\omega(0)$ - 2.5 meV (the undoped bulk value) and 5.6 meV (that was assumed in the above plots), and (3) where the phonon dispersion is considered and also where it is ignored. As context, in the phenomenological theory of Ref.~\cite{liu2}, we expect that $\lambda$ scales as $t_0^2/[\omega(0)\omega(2k_F)]$. The numerator follows trivially from Eq.~\ref{matrix}. From Eq.~\ref{matrix}, ignoring the phonon dispersion, $\lambda$ scales as $1/\omega^2(0)$ as verified in Table \ref{table3}. For the case where the phonon dispersion is kept, one might expect based on the phenomenological expression that $\lambda$ would scale as $1/\omega(0)$ instead since $\omega(2k_F)$ does not vary much with $\omega(0)$. This is indeed what we find as illustrated in Fig.~11. To explore this further, one would need to know what the actual phonon energies are at the interface, potentially exploiting techniques like the one presented in Ref.~\cite{chu}.

\begin{figure}
\centering
\includegraphics[width=1.0\columnwidth]{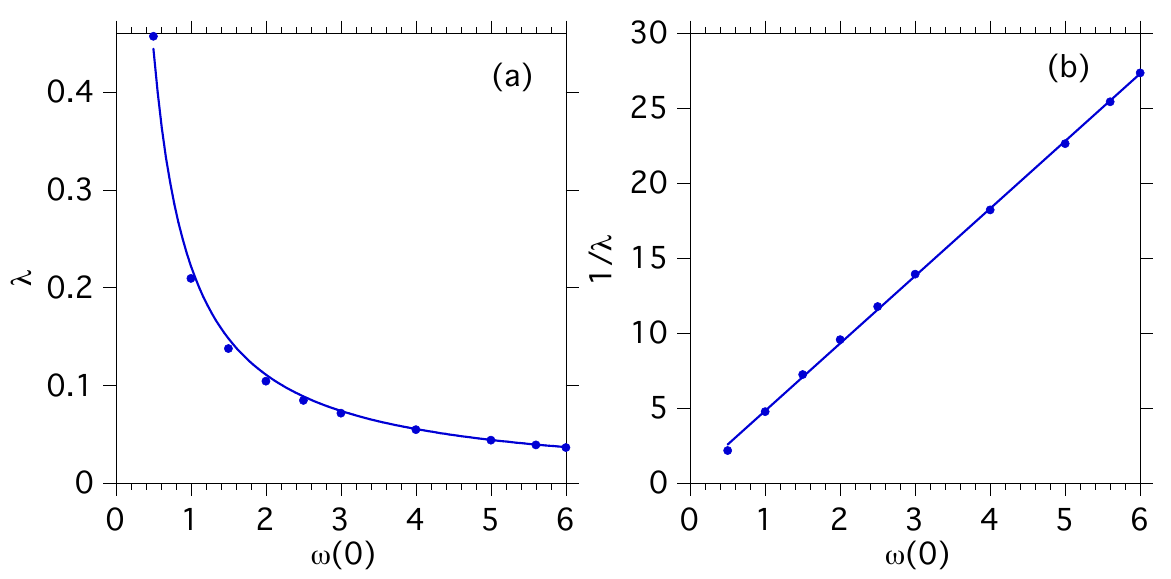}
\caption{Variation of the BCS coupling constant $\lambda$ with respect to $\omega(0)$ for the 111 bilayer case (TB model 2). The solid curves indicate that $\lambda$ scales as $1/\omega(0)$.}
\label{fig11}
\end{figure}

\begin{table}
\caption{Values of the BCS coupling constant $\lambda$ for various cases. TB1 refers to TB model 1 and TB2 to TB model 2. $\omega(0)$ (column 2) is given in units of meV. Column 3 refers to whether the phonon dispersion is taken into account, $\omega(q)$, or not, $\omega(q)=\omega(0)$.}
\begin{ruledtabular}
\begin{tabular}{llll}
 & $\omega(0)$ & $\omega$ & $\lambda$\\
\hline
111, TB1 & 5.6 & $\omega(q)$ & 0.0545\\
111, TB1 & 5.6 & $\omega(0)$ & 0.1589\\
111, TB1 & 2.5 & $\omega(q)$ & 0.1304\\
111, TB1 & 2.5 & $\omega(0)$ & 0.7974\\
\hline
111, TB2 & 5.6 & $\omega(q)$ & 0.0393\\
111, TB2 & 5.6 & $\omega(0)$ & 0.1074\\
111, TB2 & 2.5 & $\omega(q)$ & 0.0848\\
111, TB2 & 2.5 & $\omega(0)$ & 0.5392\\
\hline
001, TB1 & 5.6 & $\omega(q)$ & 0.0009\\
001, TB1 & 5.6 & $\omega(0)$ & 0.0015\\
001, TB1 & 2.5 & $\omega(q)$ & 0.0029\\
001, TB1 & 2.5 & $\omega(0)$ & 0.0075\\
\hline
001, TB2 & 5.6 & $\omega(q)$ & 0.0030\\
001, TB2 & 5.6 & $\omega(0)$ & 0.0048\\
001, TB2 & 2.5 & $\omega(q)$ & 0.0095\\
001, TB2 & 2.5 & $\omega(0)$ & 0.0238\\
\end{tabular}
\end{ruledtabular}
\label{table3}
\end{table}

\section{Discussion}
From Table \ref{table3}, one finds, as expected, that $\lambda_{111}$ is much larger than $\lambda_{001}$. But even for the 111 case, $\lambda$ is (typically) significantly less than the value of $~\sim 0.26$ that is needed to account for $T_c$ at this carrier density \cite{liu2}. Here, we mention some of the improvements that would be needed over the current theory to have a more complete theory of superconductivity. (1) coupling to the TO1 mode that is quadratic in the atomic displacements.  This intraorbital term has been argued to be important for STO \cite{vdM}. (2) coupling to the other TO1 mode (the one polarized in the plane). We note that in the presence of an electric field along the normal to the interface, this mode should be softer than the mode considered here (the one polarized along the normal) \cite{fleury}. (3) coupling to the other LO and TO modes. These couplings are known to be significant \cite{giulia1} and have been argued to be important for describing the ARPES spectral function for KTO \cite{chen-feng}. (4) a more complete model for the phonon dispersion and its dependence on carrier density. The current Vaks parameterization is an expansion around $q=0$ and does not account for the flattening of the phonons at the zone boundary. And as discussed in Ref.~\cite{liu2}, the results for $T_c$ versus $n_{2D}$ are sensitive as well to how $\omega(0)$ varies with $n_{2D}$. (5) a better estimate of the BCS cut-off $\omega_c$ which relates $\lambda$ to $T_c$. Ultimately, one will want to solve strong-coupling gap equations along the lines of Ref.~\cite{mcmillan}. (6) moving away from the simple bilayer model presented here. A more complete treatment would involve solving coupled Schroedinger-Poisson equations for the heterostructure. This would then properly describe the energies and wavefunctions of the various subbands due to the confinement potential \cite{ueno}, as well as the variation of such quantities as $\omega(q)$ with respect to $z$, the normal to the interface.

\section{Conclusion}
In this paper, we consider a microscopic theory of pairing due to exchange of the soft TO1 mode characteristic of quantum paraelectrics and applied this to a simplified version of the electronic structure at the interface of KTO heterostructures. We then use this model to explain the large variation of $T_c$ with interface orientation, as well as the variation of $T_c$ with $\omega(q)$. As a bonus, we exploit this model to look at the dependence of the gap function on both the band index and Fermi surface angle. We find this to be significant for both the 001 and 111 cases which could be studied by tunneling measurements. Note because of the assumed static Rashba term, the Kramers degeneracy of the input electronic structure in Eq.~\ref{matrix} is lifted.  Had interband terms been ignored, then $\frac{1}{2}(\Delta_1+\Delta_2)$ would be considered as the singlet order parameter, and $\frac{1}{2}(\Delta_1-\Delta_2)$ as the triplet one. But since they are present, this decomposition does not have much meaning. We also considered pairing for only the identity representation of the surface Brillouin zone. Given the forward scattering nature of the interaction, then higher angular momentum pairing becomes more relevant, though given the attractive nature of the electron-phonon interaction, we expect the identity representation to have the highest $T_c$. This would be consistent with the simplified bulk calculations shown in Ref.~\cite{maria1}.

Although we do not comment on the variation of $T_c$ with $n_{2D}$, in the phenomenological theory of Ref.~\cite{liu2}, this is sensitive to how both $\omega(0)$ and $\omega_c$ vary with $n_{2D}$.  In fact, a linear variation of $T_c$ with $n_{2D}$ is the exception rather than the rule. Typically, one finds a dome-like dependence instead \cite{liu2} which has also been seen in recent experiments \cite{eom}. In fact, one can go from a linear variation to a dome-like behavior with $n_{2D}$ by simply replacing $k_F$ in $\omega_c=\omega(2k_F)$ by its value along the $\Gamma-M$ direction in Fig.~1a (red/blue curves) instead of the $\Gamma-K$ direction that was assumed in Ref.~\cite{liu2}. This is because of the Fermi surface anisotropy. This ambiguity would be resolved if a full strong coupling calculation was done.  In that context, in the present theory we are in the adiabatic limit in that $\omega(2k_F)$ is smaller than $E_F$. But if one considers the high energy LO mode advocated in Ref.~\cite{chen-feng}, then one can be in the anti-adiabatic limit instead, which introduces new complications \cite{review}.

We now turn to the question of why no superconductivity has been seen in bulk samples, even those that are claimed to be doped. The reason this is puzzling is that the bulk is cubic, and so one has maximal orbital degeneracy in that case. Ref.~\cite{liu2} shows that $T_c$ is suppressed for the bulk since its density of states $N_{3D}$ varies as $n^{1/3}_{3D}$ and so $N_{3D}$ vanishes in the low carrier density limit. But this argument would not apply for higher carrier densities. In that context, we remind that the pairing in the heterostructure seems to be dominated by the interface, in that the mean-field $T_c$ (as defined from a Halperin-Nelson fit to the resistivity \cite{HN}) rises continuously with negative gate voltage which acts to push the carriers to the interface. This is why a complete theory of superconductivity that involves the Poisson-Schroedinger equation would be key in understanding the difference between the bulk and the heterostructure \cite{haraldsen}.

Regardless, even at the level we treat the KTO pairing problem in the present paper, we find the theory to be rich given the non-trivial nature of the TO1 mode coupling (strongly suppressed for back scattering) and the resulting dependence of $\Delta_{nk}$ on both the band index and Fermi surface angle.  Although likely beyond the precision of ARPES, information on this could be determined by tunneling.

\begin{acknowledgments} 
This work was supported by the Materials Sciences and Engineering Division, Basic Energy Sciences, Office of Science, US Dept.~of Energy.  The author acknowledges discussions with Anand Bhattacharya and Maria Gastiasoro about the topics addressed in this paper.
\end{acknowledgments}

\appendix

\section{Hamiltonian and Rashba matrix elements}
The bilayer electronic structure used for the 111 case can by found in Ref.~\cite{xiao}.
As an example, the Hamiltonian including only near-neighbor hopping terms is \cite{pablo}
\newcommand{\dg}{\dagger}
\newcommand{\pdg}{{\vphantom\dagger}}
\newcommand{\bk}{{\bf k}}
\begin{eqnarray}
\hat{\mathcal{H}}(\bm{k}) &=& \left[\xi^\pdg_\ell(\bk) d^\dg_{1,\ell\alpha}(\bk) d^\pdg_{2,\ell\alpha}(\bk)+ h.c.\right] \notag \\
&+& i \frac{\xi_{so}}{2} \varepsilon^\pdg_{\ell m n} \sigma^n_{\alpha\beta} d^\dg_{i,\ell \alpha}(\bk) d^\pdg_{i,m \beta}(\bk) \notag \\
&+& \frac{\Delta}{2} (1-\delta_{\ell,m}) d^\dg_{i,\ell \alpha}(\bk) d^\pdg_{i,m \alpha}(\bk)
\end{eqnarray}
where $\sigma$ are the spin Pauli matrices, $d$ the creation/annihilation operators for the $t_{2g}$ electrons, and
\begin{equation}
\begin{split}
 &\xi_1=-te^{ik_2 c}\left[1+e^{i\left(\frac{\sqrt{3}k_1 c}{2}-\frac{3k_2 c}{2}\right)}\right]\\
 &\xi_2=-te^{ik_2 c}\left[1+e^{-i\left(\frac{\sqrt{3}k_1 c}{2}+\frac{3k_2 c}{2}\right)}\right]\\
 &\xi_3=-2t\cos\left(\frac{\sqrt{3}k_1 c}{2}\right)e^{-i\frac{k_2 c}{2}}
\end{split}
\end{equation}
where $t$ is the near-neighbor hopping parameter (Table 1).
Here, $k_1$ is along (1,-1,0) and $k_2$ along (1,1,-2), with $c=\sqrt{2/3}a$ where $a$ is the bulk lattice constant.
For $d$, $1,2$ refers to the layer index, $l$ to the orbital index (yz, xz, xy) and $\alpha$ to the spin index. $\xi_{so}$ is the spin-orbit coupling (the same enters the $t_{2g}-e_g$ coupling terms discussed in the main text, but with different prefactors \cite{shanavas2}).  $\Delta$ is the trigonal distortion, which is ignored here since it is known from ARPES \cite{bruno} to be small and we are in the limit that $E_F >> \Delta$. The longer range hopping expressions can be found in Ref.~\cite{xiao}.  The bilayer 001 hoppings are trivial to derive since they just involves the cubic expressions with the dependence on $k_z$ suppressed.

For completeness, we also list the Rashba matrix elements as presented in Ref.~\cite{maria1} for the 001 case:
\begin{eqnarray}
t_0^{yz} = -2 i t_0 \sin{(k_ya)} \sigma_0, & & t_0^{zx} = +2 i t_0 \sin{(k_xa)} \sigma_0 \nonumber \\
t_A^{xx} = +2 t_A \sin{(k_ya)} \sigma_x, & & t_A^{yy} = -2 t_A \sin{(k_xa)} \sigma_y \nonumber \\
t_B^{xy} = -2 t_B \sin{(k_ya)} \sigma_y, & & t_B^{xy} = +2 t_B \sin{(k_xa)} \sigma_x \nonumber \\
t_C^{zx} = -2 t_C \sin{(k_ya)} \sigma_z, & & t_C^{yz} = +2 t_B \sin{(k_xa)} \sigma_z \nonumber
\end{eqnarray}
Here, the $t_{2g}$ orbitals are mapped as follows: $yz$ to $x$, $xz$ to $y$ and $xy$ to $z$, with $a$ the bulk lattice constant.
Note that $t_0$ is the spin-independent Rashba coupling denoted as $\alpha$ in Ref.~\cite{shanavas}, whereas the spin-dependent terms correspond to $\beta$ and $\gamma$ of Ref.~\cite{shanavas}. The values of the various $t_i$ are listed in Table \ref{table2}.

For the 111 case, the remaining matrix elements can be determined by permuting $x,y,z$ including the index for $\sigma$. The momentum dependent terms $\sin(k_xa)$, etc., are replaced by the corresponding (complex) quantities of the bilayer 111 surface Brillouin zone as in Refs.~\cite{xiao,pablo} and these quantities must be permuted as well. Note that the values of $t_i$ in Table \ref{table2} need to be scaled down by a factor of $1/\sqrt{3}$ for the 111 case since the phonon polarization is $n_{ph}=(1,1,1)/\sqrt{3}$.

\section{$\omega(0)$ estimate}
This follows the argumentation of Ref.~\cite{liu2}. The presence of a finite $n_{2D}$ leads to screening and so a reduction of the dielectric constant $\epsilon$, which in turn enhances $\omega_0$ via the Lyddane-Sachs-Teller relation $\omega^2(0) \propto 1/\epsilon$, as has been seen in experiments on bulk samples (the longitudinal mode remains at much higher frequencies). To proceed, we first need to know the electric field dependence of $\epsilon$ which we derive from field-dependent Raman studies of KTO \cite{fleury}. From that work, we take the relative dielectric function to be $\epsilon(E) = 4500/(1+bE)$ where $b=8 \times 10^7$ ($E$ in V/m). To determined the relevant `average' field $F$, we follow Ref.~\cite{ueno-STO} where a triangular confining potential was assumed along $z$ (the normal to the interface) with the dielectric function integrated in field up to $F$. That is $en_{2D} = 2 \int_0^F \epsilon_0 \epsilon(E) dE$ where $\epsilon_0$=8.854 pF/m.  For $n_{2D}=10^{14}cm^{-2}$, $\epsilon(F)$ is reduced from 4500 to 897, resulting in an $\omega_0$=5.6 meV.

\bibliography{main}

\end{document}